\documentclass[twocolumn,prb,showpacs,preprintnumbers,amsmath,amssymb]{revtex4}

\usepackage{graphicx}
\usepackage{dcolumn}
\usepackage{bm}

\begin{document}

\title{Magnetooptic enhancement and magnetic properties in Fe antidot films with hexagonal symmetry }

\author{E. Th. Papaioannou}
\email{vangelis@fysik.uu.se}
\author{V. Kapaklis}
\affiliation{Department of Physics and Materials Science, Uppsala University, 75121 Uppsala, Sweden}
\author{P. Patoka}
\affiliation{Helmholtz-Zentrum Berlin f\"ur Materialien und Energie GmbH, 14109 Berlin, Germany}
\author{M. Giersig}
\author{P. Fumagalli}
\affiliation{Institut f\"ur Experimentalphysik, Freie Universit\"at Berlin, 14195 Berlin, Germany}
\author{A. Garcia-Martin}
\author{E. Ferreiro-Vila}
\affiliation{Instituto de Microelectr\`onica de Madrid (IMM-CNM-CSIC), 28760 Madrid, Spain}
\author{G. Ctistis}
\email{g.ctistis@utwente.nl}
\affiliation {FOM Institute for Atomic and Molecular Physics (AMOLF), Center for Nanophotonics, 1098 XG Amsterdam, The Netherlands}
\affiliation{Complex Photonic Systems (COPS), MESA+ Institute for Nanotechnology, University of Twente, The Netherlands}
 
\date{\today}

\begin{abstract}
The magnetooptic and magnetic properties of hexagonal arrays of holes in optically thin iron films are presented. We analyze their dependence on the hole radius and compare the results to a continuous iron film of same thickness. We observe a large enhancement of the magnetooptic Kerr rotation with respect to that of the continuous film, at frequecies where surface plasmon excitations are expected. The spectral position of the Kerr maxima can be tuned by the size and the distance between the holes. Additional simulations are in very good agreement with the experiment and thus confirm the effect of the surface plasmons on the Kerr rotation. The altering of the magnetic properties by the hole array is also visible in the hysteretic behavior of the sample where a significant hardening is observed.

\end{abstract}

                             

\maketitle

\section{\label{sec:level1}Introduction}
Patterning holes into ferromagnetic thin films (antidots) is an effective way to engineer their magnetic properties. The presence of antidots alters the demagnetization fields of the structures. At the same time the holes are pinning sites for domain walls. These two parameters have been shown to influence the coercivities and remanences \cite{Ctistis2009, adeyeye97}, anisotropies \cite{yu:6322, vavassori:7992}, and switching characteristics \cite{wang2003, heyderman2006}.
In parallel, the antidot structures with noble metals have been studied for their optical properties after the pioneering work of Ebbesen \textit{et al.} \cite{Ebbesen1998} and the discovery of extraordinary optical transmission of light through these subwavelength structures at certain resonant frequencies or angle of incidence. The investigations have not included the presence of a magnetic field since it is known that for plasmonic metals the influence of a magnetic field is very low and thus the use of very high magnetic fields would be necessary to observe interactions. However, by using ferromangetic materials the interactions with a magnetic field is much stronger than for the noble metals that would in principle lead to measurable magneto-plasmonic interactions, even in the presence of the absorptive losses for the plasmons. Indeed, the interaction of light  with the ferromagnetic nanoscale arrays of holes in an applied magnetic field has shown exciting optical and magnetooptic properties. In particular,  magneto-plasmonic interactions were observed by Ctistis \textit{et al.} \cite{Ctistis2009}, where an extraordinary magnetooptic response of antidot hexagonal arrays of Co was revealed, and by Gonzalez-Diaz \textit{et al.} \cite {Diaz07} who reported an enhancement of the magnetooptic response for Ni nanowires embedded in an alumina matrix due to the propagation of plasmons in the nanowires. Furthermore recently, predictions of a signiÞcant enhancement of the magneto-optic transverse Kerr effect were reported for the case of a magnetic/noble metal film perforated with subwavelength slit arrays \cite{Belotelov09}. 

In this work, we have succesfully fabricated subwavelength hole arrays in optically thin Fe films with different hole sizes. We present a detailed study on the size dependencies of the magnetic and magnetooptic properties.

\section{\label{sec:level2}Experimental details}
\begin{figure}
\includegraphics{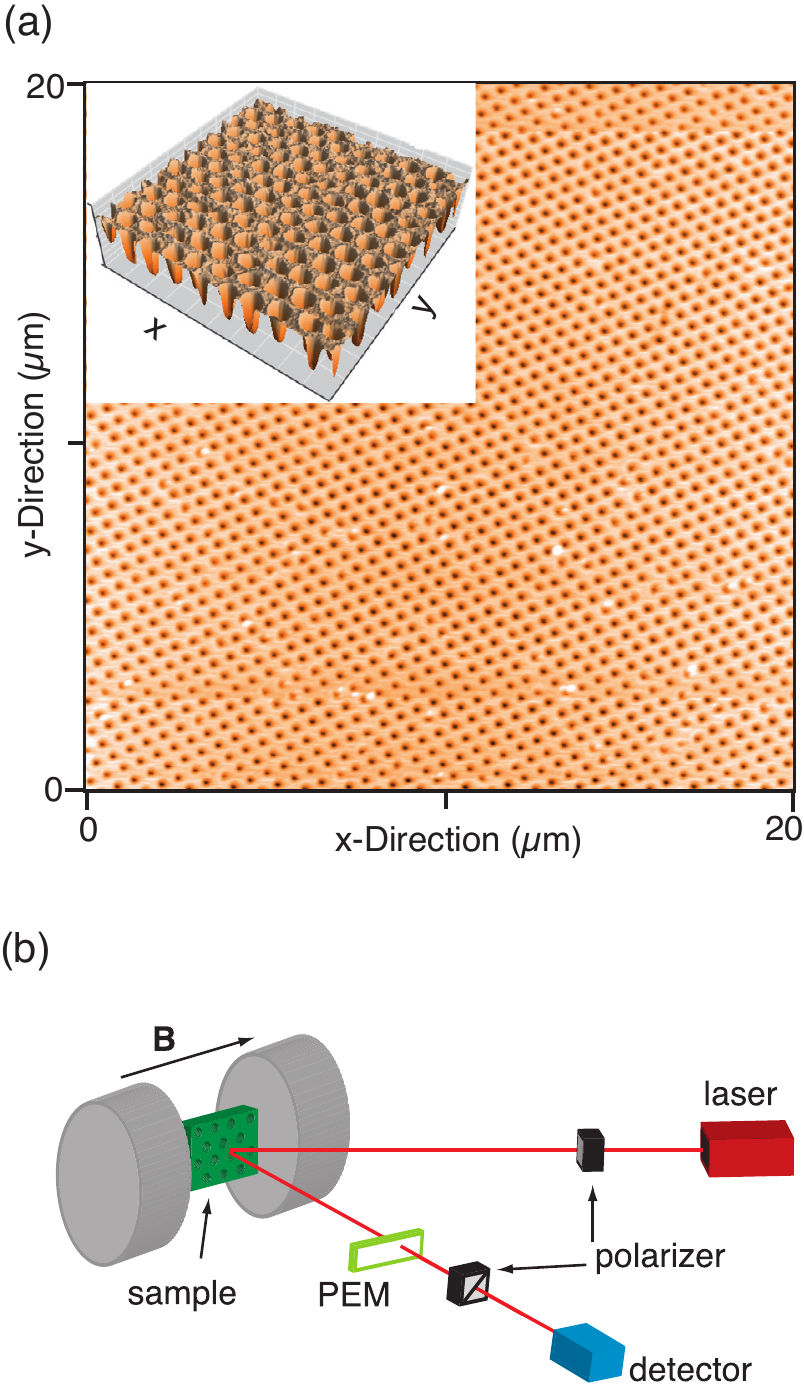}
\caption{\label{fig:fig1}(color online) (a) Atomic force micrograph of a $100\ nm$ thick Fe film with hexagonal arrays of holes. The pitch size of the array is $a=470\ nm$ and the hole size is $d=248\ nm$. A large defect-free area is visible. The inset shows a 3D representation of the atomic force micrograph.
(b) Schematic of the experimental setup for the longitudinal Kerr measurements. The angle of incidence of the light is $24^{\circ}$. } 
\end{figure}

The Fe antidot arrays were prepared on a Si(111) substrate using self-assembly nanosphere lithography with polystyrene (PS) spheres. For a more detailed description of the process we refer to Ref.\cite{Giersig2004} and will give only a short summary here.
We used monodisperse PS spheres with a diameter of $470\ nm$ as a mask template. After the formation of a hexagonal closed-packed structure with the PS-spheres, the diameter of the spheres has been shrunk by means of reactive ion-etching (RIE). The control of the etching parameters enables us to control the diameter of the PS-spheres, here $248$ and $297\ nm$, respectively.
The thus obtained structure served as a mask for the evaporation of the $100\ nm$ thick iron films by means of molecular-beam epitaxy and a base pressure of $10^{-7}\ mbar$. The used evaporation conditions lead to a polycrystalline structure of the iron films. In order to obtain a smooth and continuous Fe film, a seed layer of $2\ nm$ Ti was deposited prior the Fe evaporation. To prevent subsequent oxidation of the surface a capping layer of $2\ nm$ gold was deposited on top. 
Thereafter, a chemical treatment dissolved the PS-spheres leaving the metal film with a hexagonal array of holes behind [see Fig. \ref{fig:fig1}(a)]. Due to this treatment and the measurements under ambient conditions, a final oxidation of the Fe films can not be excluded.
To characterize the topography as well as the micromagnetic behavior of the samples, an atomic force microscope (Nanosurf Mobile S) operating in tapping mode was used. By measuring the phase shift of the cantilever oscillation we were able to observe the magnetic interaction of the cobalt coated cantilever tip with the underlying sample (oscillating frequency $75\ kHz$, spring constant $2.8\ N/m$). An AFM image of the 3-dimensional reconstruction of the patterned Fe film with holes diameter $248\ nm$ is presented in the inset of Fig. \ref{fig:fig1}(a). The MFM measurements were performed at room temperature without applied external field implying that the remanent magnetic state of the sample is studied. Additionally, we performed micromagnetic simulations using an object-oriented micromagnetic framework (oommf) from NIST \cite{oommf} to explain micromagnetic distributions to the spin orientation imaged with the magnetic-force microscope.

Furthermore, the macroscopic optical and magnetic properties of the films were investigated using a magnetooptic Kerr spectrometer in the longitudinal and polar configuration under ambient conditions. A schematic of the longitudinal Kerr setup is shown in Fig. \ref{fig:fig1}(b).
The longitudinal Kerr magnetometer, based on the use of a photoelastic modulator (PEM) operating at $50\ kHz$, allowed the simultaneuous meauserement of both Kerr rotation and ellipticity. The incident polarization set by the first polarizer, corresponds to s-polarized light. After the sample, the beam passes through the modulator and an analyser. The PEM retardation axis was parallel to the plane of incidence. The analyzer is oriented at $45^{\circ}$ with respect to the PEM retardation axis. The measurements were performed at a photon energy of $1.878\ eV$ at an angle of incidence of $24^{\circ}$ with respect to the sample surface normal. 
The polar Kerr magnetometer, based on the use of a Faraday modulator \cite{papaioannou2007}, was used to record polar Kerr hysteresis loops at selected energies at the maximum magnetic field of $1.64\ T$. Furthermore, polar-MOKE spectroscopy at room temperature (RT) was performed with an applied magnetic field of $1.64\ T$, at an angle of incidence of $5^{\circ}$ with respect to the sample surface normal and at photon energies between $1.5 $ and $4.3\ eV$. 

\section{\label{sec:level2}Results and Discussion}

\begin{figure}
\includegraphics{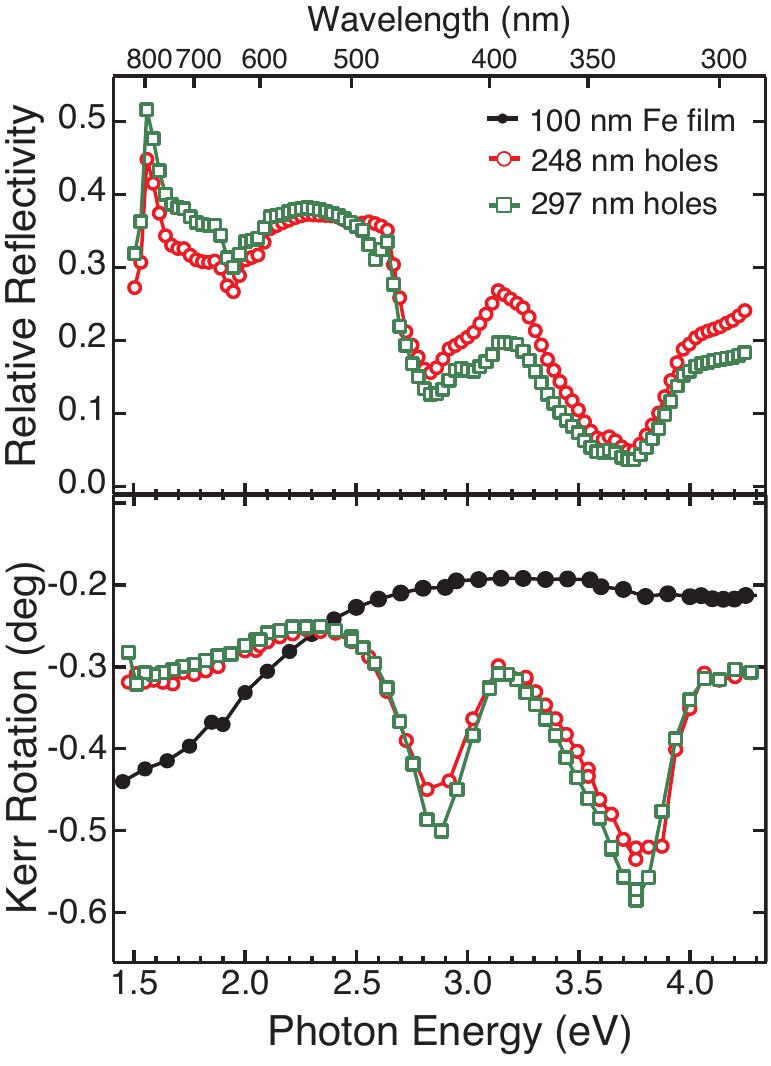}
\caption{\label{fig:fig2} (color online) Relative reflectivity (top) and magnetooptic polar Kerr (bottom) measurements for two different Fe hole diameters with holes diameter of 248 nm (open circles) and 297 nm (open squares), respectively. The Kerr spectrum of a continuous Fe film (filled circles) is shown as reference. The Kerr spectra are recorded at the saturation state of the samples at magnetic field of $1.64\ T$.  The magnetooptic response of the antidot samples is strongly affected and enhanced at $\sim2.8\ eV$, and at $\sim3.8\ eV$ as a direct response to surface plasmon excitation.}
\end{figure}

Typical reflectivity spectra are plotted together  with polar magnetooptic spectra in Fig. \ref{fig:fig2}. The reflectivity spectra shown in Fig. \ref{fig:fig2} (top graph) are measured for two different hole arrays with $248\ nm$ (circles) and $297\ nm$ (squares) hole diameter, respectively. The spectra show similar spectral behavior. The main features are the minima in reflectivity at $\sim2.8\ eV$ and at $\sim3.8\ eV$. These minima are a result of the resonant coupling of light to surface plasmon (SP) excitations of both interfaces of the Fe film perforated with a hexagonal array of holes with a lattice constant of $a=470\ nm$.
Figure \ref{fig:fig2} (bottom graph) shows the magnetooptic spectra of the two hole array samples. Additionally, the spectrum of a Fe film of same thickness as for the hole arrays is shown. In the low energy regime, the Kerr rotation is smaller than the continuous film, which is expected, if we consider that we have a smaller amount of magnetooptically active material in the hole arrays. Nevertheless, above $2.5\ eV$ a very strong  enhancement of the Kerr rotation is observed for both hole arrays with two maxima around $2.8\ eV$ and $3.8\ eV$. The maximum Kerr rotation values at these energies are nearly two and three times bigger than the values of the continuous film. It is worth to notice that the spectacular enhancement is visible at the same energies as the features in the reflectivity spectra, mirroring the fact that surface-plasmon excitation of the array influences the magnetooptic properties of the Fe film. Above $4\ eV$,  Kerr rotation is decreasing, however it is still much bigger than the one of the continuous film. The results are similar to earlier measurements on Co hole arrays \cite {Ctistis2009}. In both cases we have a strong enhancement of the polar Kerr values which is supported by surface-plasmon resonances at specific energies.

\begin{figure}
\includegraphics{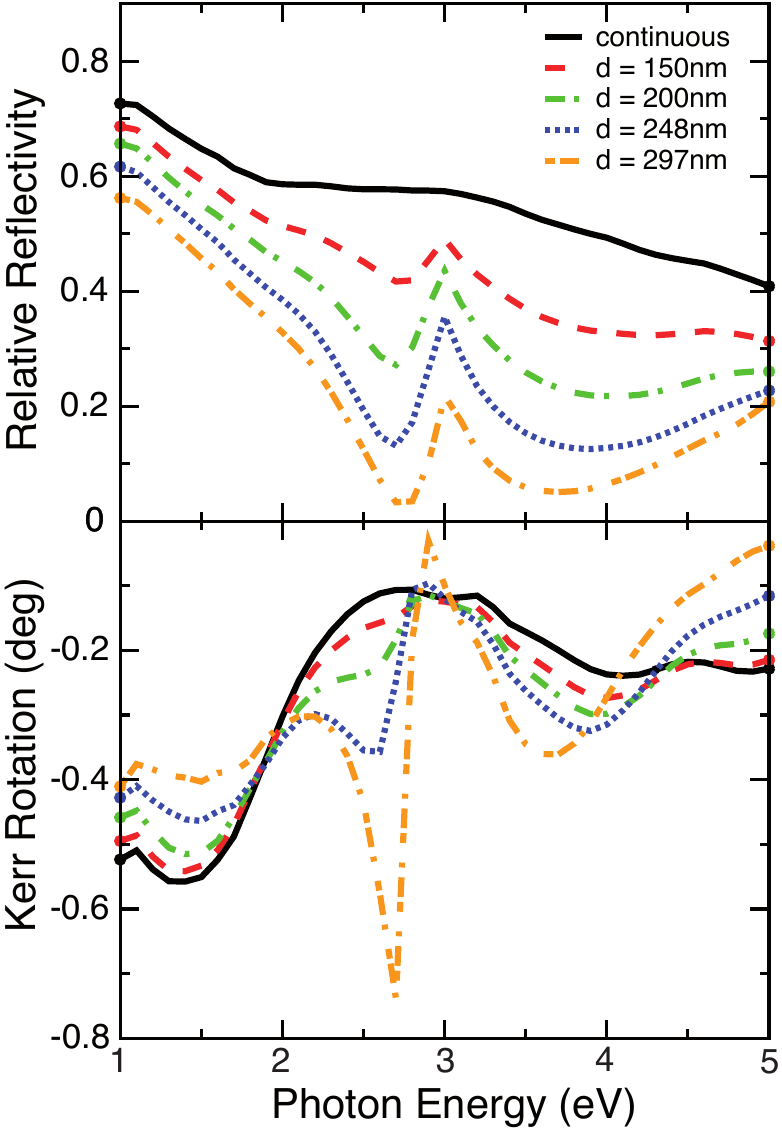}
\caption{\label{fig:fig3} (color online) Calculated reflectivity and polar Kerr rotation versus photon energy. The array parameters for the hole arrays are as follows: $a=470\ nm$ as pitch size for all arrays, $d=150\ nm$, $d=200\ nm$, $d=248\ nm$ and $d=297\ nm$. The two bigger hole diameter correspond to the experimental spectra shown in Fig. \ref{fig:fig2}. The spectrum for the Fe film is also shown. A continuous enhancement of the polar Kerr rotation as the reflectivity decreases is observed as the hole diameter increases.}
\end{figure}

For a more thorough understanding of the underlying effects, a theoretical approach has been performed. The calculations employed a scattering matrix method specifically adapted to consider magnetooptic effects \cite{garcia05}. In Fig. \ref{fig:fig3} calculated spectra for the reflectivity (top graph) and the polar Kerr rotation (bottom graph) of the corresponding two hole arrays of Fe and the continuous film are presented. Furthermore, in the same graph, we have simulated two more samples with hole diameters of $150\ nm$ and $200\ nm$ in order to reveal how the diameter $d$ rules the behavior of such arrays.

The simulations are in good qualitative agreement with the experimental results and provide guidelines to artificially control the optical and magnetooptic properties of the antidots. The calculated reflectivity curves exhibit similar features for all samples. The first minimum is almost constant at $2.8\ eV$ indicative of the fact that the optical response depends on the interhole separation $a=470\ nm$, that is the same for all the samples. The interaction of light with the hexagonal periodicity of hole arrays leads to enhancement of plasma oscillations of the electrons (minimum in reflectivity) at resonant frequencies. The second minimum is much broader and shifts slightly towards lower energies as the diameter of the holes increases. The features of the calculated curves agree quite well with the experimental reflectivity of the two samples at the positions of the two minima at $2.8\ eV$, and $3.8\ eV$.  

The enhancement of plasmon oscillations at specific energies is coupled to the magnetooptic response. The simulations of the Kerr spectra reproduce the experimental enhancement of the magnetooptic activity at the end of the visible region and in the UV. The energy positions of the maxima agree well with the experiment. Even more, the simulations reveal clearly the behavior of the antidots with different diameters. The Kerr response is enhanced as we go to higher diameters and the reflectivity decreases. The presence of bigger holes brings the nanoholes closer and enables the excitation of surface plasmons to interact more sufficiently with the adjacent nanoholes. As a result  the enhancement of the Kerr effect is  maximized at the resonant frequency of $2.8\ eV$ for the sample with $d = 297\ nm$. At the same time, by increasing the diameter a shift to lower energies for the second Kerr maximum at energies above $3.5\ eV$ is observed.

The calculated absolute rotation values are different than the experimentally measured. The peak at $2.8\ eV$ is more pronounced in the simulation than the one at $3.8\ eV$. The difference could be attributed to an insufficient description of the optical constants of the materials, especially inside the holes. It is expected that the presence of an oxide Fe layer at the side walls of the holes will modify the refractive index $n$ of the material and consequently changing the plasmonic characteristics responsible for the magnetooptic enhancement. Recently, Gonzalez-Diaz \textit{et al.} \cite{garcia09} reveal the role of the increase of the refractive index of a material that fills the pores in the magnetooptic response by observing a red-shift in the Kerr rotation maxima.

\begin{figure}
\includegraphics{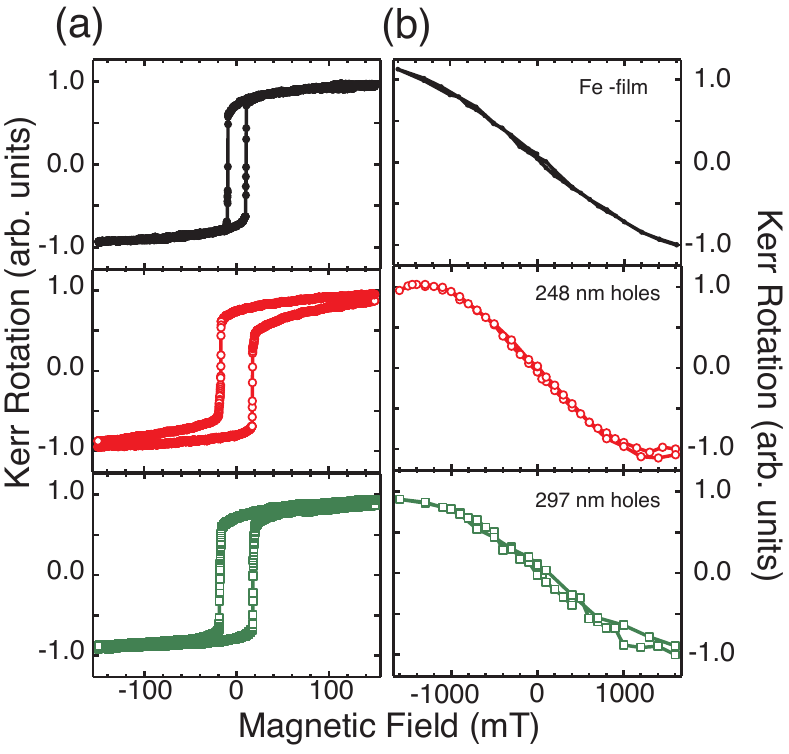}
\caption{\label{fig:fig4}(color online) Longitudinal (a) and polar Kerr (b) hysteresis loops for a Fe thin film (closed circles) and for two antidot Fe samples with $d = 248\ nm$ (open circles) and $297\ nm$ (open squares) hole diameter. The lattice constant of the antidot array is $a=470\ nm$ for both samples. The y-axis is normalized to unity for a better comparison. The magnetic behavior is dominated by the size of the holes. Higher coercivity and the appearance of out-of-plane magnetization components with increasing hole size is visible.}
\end{figure}

To investigate the magnetic behavior closer, hysteresis loops both in longitudinal and polar configuration were measured. Figure \ref{fig:fig4} shows the results for a Fe film (top row), a $248\ nm$ hole array (middle row), and a $297\ nm$ hole array (bottom row), respectively. The hysteresis curve for the continuous film, taken in longitudinal configurationÊ[Fig. \ref{fig:fig4}(a)], confirms that the easy axis of the magnetization lies in the film plane due to the shape anisotropy of the Fe film. 
Keeping the applied magnetic film in-plane but in different directions of the 2D unit cell, covering thereby the full $360^{\circ}$ range, enables us to determine that the sample does not show any in-plane anisotropy, as expected for a polycrystalline film. Since the material itself does not possess an intrinsic magnetic anisotropy,  the different hysteresis loops that we observe for the antidot films indicate the dominant role of the size and the arrangement of the holes in the reversal magnetic behavior. The two patterned samples exhibit a larger coercivity ($H_{C}\approx20\ mT$), which is twice as large compared to the continuous film. The trend of magnetic hardening for the hole arrays can be attributed to the presence of the holes. 

In particular, the holes introduce large areas of air-metal boundaries, only around $70\%$ of the surface is covered with material. As a consequence these boundaries modify the demagnetizing field distribution in the film. At the same time they serve as domain wall pinning sites. Even more, the shape of the hysteresis curves is markedly modified by the presence of the nonmagnetic holes, they are more squared in shape.

Although the hexagonal arrangement of the holes should introduce a threefold in-plane anisotropy as observed in micromagnetic simulations \cite{wang06}, measurements of hysteresis loops in various in-plane magnetic field directions do not show any changes, neither in coercive field ($H_{C}$) nor in saturation field ($H_{sat}$). The absence of a hard and easy in-plane axis can be explained by the fact that the illumination spot during the experiment is not focussed and covers an area ($\sim1\ mm^{2}$), which probes more than one structural domain. Therefore, we average over all possible orientations breaking  thereby the in-plane shape anisotropy. 

Figure \ref{fig:fig4}(b) shows hysteresis measurements in the polar configuration. As can be seen from the graphs, all samples have their hard magnetization axis out of plane. For the reference sample, the hysteresis loop shows a typical hard-axis behavior with no remanence and a saturation field that we have hardly reached with our setup of $B=1.64\ T$. The hole arrays, however, have modified the hysteresis loops. By increasing the hole size the saturation field decreases strongly and reaches the value of $H_{sat}=1\ T$ for the sample with $297\ nm$ hole diameter. Simultaneously, a small hysteresis and remanence appear. The local dipolar fields introduced by the hole edges give rise to the out of plane magnetization components in competition with the intrinsic in plane anisotropy of the samples.

\begin{figure}
\includegraphics{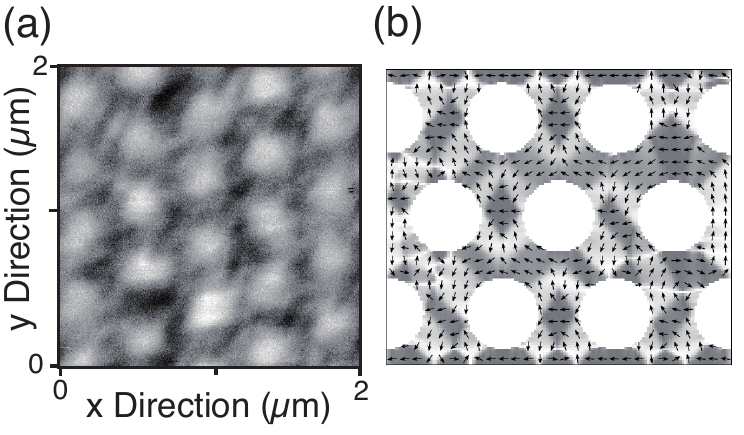}
\caption{\label{fig:fig5} (a) Magnetic-force micrograph of a nanohole array ($d=248\ nm$) with a $100\ nm$ thick Fe film without applied external magnetic field. The color scale describes different magnetization orientations. (b) Micromagnetic simulation of the same structure as in (a) in remanence. The spin orientation is denoted by the arrows and the color scale, showing mainly an in-plane orientation affected by the holes.}
\end{figure}

Magnetic force microscopy was used to visualize the formation of magnetic domains in such systems in the remanent state. Figure \ref{fig:fig5}(a) shows a magnetic force micrograph of the sample with the $248\ nm$ holes. Bright and dark regions are distributed among the holes without any correlation to the simultaneously recorded topography (not shown here). The MFM technique is based on the interactions of the tip with the magnetic charges in the sample. This interaction gives information about the stray field of the sample. The stray field of the patterned sample is strongly influenced by the presence of the holes since the magnetic field lines can close through the holes. The strong contrast can be attributed to small out-of-plane components (see Fig. \ref{fig:fig4}(b)), creating magnetic poles that give rise to dark or bright regions. Similar domain configuration was observed in a Ni film of $55\ nm$ thickness but for much smaller hole diameter of $50\ nm$ \cite{jaafar07}. Even though the contrast is not so large in between neighboring holes, one can recognize upon six darker regions (different domain configurations) around all the holes  separated by lighter areas.

Figure \ref{fig:fig5}(b) shows micromagnetic simulations performed in remanence. 
The boundary conditions of the simulation are open and the calculated cell is displayed in Figure \ref{fig:fig5}(b). The dimensions in the simulation are identical with that of the real structure and are given to be $1.4 \mu m \times 1.2 \mu m$ and the cell size of $10\ nm$ was used in order to reduce computation time. We also calculated the magnetic configuration for a larger structure with more than the displayed structural unit cell but found no differences between the calculations.
The magnetic history of the sample in the simulation is set to be the same as in the experiment during the MFM measurements and is as follows: Starting with a random magnetization we first magnetize the sample in the negative x-direction with an applied field of $B=1\ T$ and then turn off the field and leave the magnetization relaxing. The parameters used for the calculations are: saturation magnetization $M_{S}=1.7\times10^6\ Am^{-1}$, exchange constant $A=21\times10^{-12}\ Jm^{-1}$, and a cubic anisotropy with an anisotropy constant of $K_{1}=48\times10^3\ Jm^{-3}$.  As convergence criterion the misalignment between magnetization and effective field was used and set to be lower than $10^{-5}$ in each computation cell. The film thickness was $100\ nm$.
The arrows in the picture denote the in-plane spin orientation of the computation cells. It is visible that the holes affect the spin orientation. In particular, the remanent spin configuration can generally be divided in different arrangements. One group is along the x-direction ($0^{\circ}$ with the x-axis). Domains are pinned along this direction and they are placed in the central region among adjacent holes. Different groups of spin configurations are formed having an angle $\pm30^{\circ}$, $\pm60^{\circ}$, $\pm90^{\circ}$ to the x-axis. These configurations smoothly circle around the holes as a result of minimizing the total energy between two competing terms: the magnetostatic and the exchange energy across the domain walls. We can see that the simulation is in qualitative agreement with the experiment.
The revealed correlation between the formation of domains and the periodic structure results in pinning effects, which justify the magnetic hardening observed in the MOKE loops in Fig. \ref{fig:fig4}.

\section{Conclusions}

The magnetic and magnetooptic properties of hexagonal arrays of holes in thin Fe films were presented. We analyzed the dependence of the magnetic and magnetooptic properties on the hole size and we compared them with a similar continuous Fe film. Extraordinary enhancement of the magnetooptic Kerr rotation is observed, which is related to the surface plasmon resonances and the hole diameter. The very good agreement with theoretical simulations gives us the ability to fully control of the properties and apply the structures for technological applications. Keeping constant the interhole distance $a$ the maximum magnetooptic  enhancement can be tuned by increasing the hole diameter. The magnetic characterization revealed the magnetic hardening, and the presence of out of plane magnetization components that give rise to different domain configurations around the holes.

\begin{acknowledgments}
E. Th. P. acknowledges the financial support from the Icelandic Science Foundation and the Swedish Foundation for International Cooperation in Research and Higher Education (STINT). M. G. thanks the Helmholtz-Zentrum Berlin for financial support. A.G.-M. and E. F.-V. acknowledge financial support from the EU Project NMP3-SL-2008-214107-Nanomagma and from the Spanish MICINN (Consolider 2010 ref. CSD2008-00023-Funcoat and MAT2008-06765-C02-01/NAN). E.F.-V also acknowledges financial support from the CSIC via the JAE-Pre program. The authors acknowledge also the Knut and Alice Wallenberg Foundation.
\end{acknowledgments}

\end{document}